\documentclass[aps,prl,twocolumn,superscriptaddress,showpacs,preprintnumbers,amsmath,amssymb]{revtex4-1}

\usepackage{graphicx} 
\usepackage{dcolumn}  

\input{babarsym.tex}

\graphicspath{{ps}}

\begin{document}

\title{ \quad\\[1.0cm] First Observation of {\boldmath\CP} Violation in {\boldmath$\Bzb \to D^{(*)}_{\CP} h^{0}$} Decays by a Combined Time-Dependent Analysis of \babar\ and Belle Data}

\newcommand{\bbr}{${^\mathrm{A}}$}
\newcommand{\bel}{${^\mathrm{B}}$}
\noaffiliation
\affiliation{Laboratoire d'Annecy-le-Vieux de Physique des Particules (LAPP), Universit\'e de Savoie, CNRS/IN2P3,  F-74941 Annecy-Le-Vieux, France}
\affiliation{Universitat de Barcelona, Facultat de Fisica, Departament ECM, E-08028 Barcelona, Spain }
\affiliation{INFN Sezione di Bari$^{\mathrm{a}}$; Dipartimento di Fisica, Universit\`a di Bari$^{\mathrm{b}}$, I-70126 Bari, Italy }
\affiliation{University of the Basque Country UPV/EHU, 48080 Bilbao, Spain }
\affiliation{Beihang University, Beijing 100191, China }
\affiliation{University of Bergen, Institute of Physics, N-5007 Bergen, Norway }
\affiliation{Lawrence Berkeley National Laboratory and University of California, Berkeley, California 94720, USA }
\affiliation{Ruhr Universit\"at Bochum, Institut f\"ur Experimentalphysik 1, D-44780 Bochum, Germany }
\affiliation{University of Bonn, 53115 Bonn, Germany }
\affiliation{University of British Columbia, Vancouver, British Columbia, Canada V6T 1Z1 }
\affiliation{Brunel University, Uxbridge, Middlesex UB8 3PH, United Kingdom }
\affiliation{Budker Institute of Nuclear Physics SB RAS, Novosibirsk 630090, Russian Federation }
\affiliation{Novosibirsk State University, Novosibirsk 630090, Russian Federation }
\affiliation{Novosibirsk State Technical University, Novosibirsk 630092, Russian Federation }
\affiliation{University of California at Irvine, Irvine, California 92697, USA }
\affiliation{University of California at Riverside, Riverside, California 92521, USA }
\affiliation{University of California at Santa Barbara, Santa Barbara, California 93106, USA }
\affiliation{University of California at Santa Cruz, Institute for Particle Physics, Santa Cruz, California 95064, USA }
\affiliation{California Institute of Technology, Pasadena, California 91125, USA }
\affiliation{Faculty of Mathematics and Physics, Charles University, 121 16 Prague, Czech Republic }
\affiliation{Chonnam National University, Kwangju 660-701, South Korea }
\affiliation{University of Cincinnati, Cincinnati, Ohio 45221, USA }
\affiliation{University of Colorado, Boulder, Colorado 80309, USA }
\affiliation{Colorado State University, Fort Collins, Colorado 80523, USA }
\affiliation{Deutsches Elektronen--Synchrotron, 22607 Hamburg, Germany }
\affiliation{Technische Universit\"at Dortmund, Fakult\"at Physik, D-44221 Dortmund, Germany }
\affiliation{Laboratoire Leprince-Ringuet, Ecole Polytechnique, CNRS/IN2P3, F-91128 Palaiseau, France }
\affiliation{University of Edinburgh, Edinburgh EH9 3JZ, United Kingdom }
\affiliation{INFN Sezione di Ferrara$^{\mathrm{a}}$; Dipartimento di Fisica e Scienze della Terra, Universit\`a di Ferrara$^{\mathrm{b}}$, I-44122 Ferrara, Italy }
\affiliation{INFN Laboratori Nazionali di Frascati, I-00044 Frascati, Italy }
\affiliation{INFN Sezione di Genova$^{\mathrm{a}}$; Dipartimento di Fisica, Universit\`a di Genova$^{\mathrm{b}}$, I-16146 Genova, Italy  }
\affiliation{Justus-Liebig-Universit\"at Gie\ss{}en, 35392 Gie\ss{}en, Germany }
\affiliation{Gifu University, Gifu 501-1193, Japan }
\affiliation{SOKENDAI (The Graduate University for Advanced Studies), Hayama 240-0193, Japan }
\affiliation{Indian Institute of Technology Guwahati, Guwahati, Assam, 781 039, India }
\affiliation{Gyeongsang National University, Chinju 660-701, South Korea }
\affiliation{Hanyang University, Seoul 133-791, South Korea }
\affiliation{University of Hawaii, Honolulu, Hawaii 96822, USA }
\affiliation{Universit\"at Heidelberg, Physikalisches Institut, D-69120 Heidelberg, Germany }
\affiliation{High Energy Accelerator Research Organization (KEK), Tsukuba 305-0801, Japan }
\affiliation{Humboldt-Universit\"at zu Berlin, Institut f\"ur Physik, D-12489 Berlin, Germany }
\affiliation{IKERBASQUE, Basque Foundation for Science, 48013 Bilbao, Spain }
\affiliation{Indian Institute of Technology Bhubaneswar, Satya Nagar 751007, India }
\affiliation{Indian Institute of Technology Madras, Chennai 600036, India }
\affiliation{Institute of High Energy Physics, Chinese Academy of Sciences, Beijing 100049, China }
\affiliation{Institute of High Energy Physics, Vienna 1050, Austria }
\affiliation{Institute for High Energy Physics, Protvino 142281, Russian Federation }
\affiliation{Institute for Theoretical and Experimental Physics, Moscow 117218, Russian Federation }
\affiliation{University of Iowa, Iowa City, Iowa 52242, USA }
\affiliation{Iowa State University, Ames, Iowa 50011-3160, USA }
\affiliation{J. Stefan Institute, 1000 Ljubljana, Slovenia }
\affiliation{Physics Department, Jazan University, Jazan 22822, Kingdom of Saudi Arabia }
\affiliation{Johns Hopkins University, Baltimore, Maryland 21218, USA }
\affiliation{Kanagawa University, Yokohama 221-8686, Japan }
\affiliation{Institut f\"ur Experimentelle Kernphysik, Karlsruher Institut f\"ur Technologie, 76131 Karlsruhe, Germany }
\affiliation{Kennesaw State University, Kennesaw GA 30144, USA }
\affiliation{King Abdulaziz City for Science and Technology, Riyadh 11442, Kingdom of Saudi Arabia }
\affiliation{Department of Physics, Faculty of Science, King Abdulaziz University, Jeddah 21589, Kingdom of Saudi Arabia }
\affiliation{Korea Institute of Science and Technology Information, Daejeon 305-806, South Korea }
\affiliation{Korea University, Seoul 136-713, South Korea }
\affiliation{Kyungpook National University, Daegu 702-701, South Korea }
\affiliation{Laboratoire de l'Acc\'el\'erateur Lin\'eaire, IN2P3/CNRS et Universit\'e Paris-Sud 11, Centre Scientifique d'Orsay, F-91898 Orsay Cedex, France }
\affiliation{\'Ecole Polytechnique F\'ed\'erale de Lausanne (EPFL), Lausanne 1015, Switzerland }
\affiliation{Lawrence Livermore National Laboratory, Livermore, California 94550, USA }
\affiliation{University of Liverpool, Liverpool L69 7ZE, United Kingdom }
\affiliation{Faculty of Mathematics and Physics, University of Ljubljana, 1000 Ljubljana, Slovenia }
\affiliation{Queen Mary, University of London, London, E1 4NS, United Kingdom }
\affiliation{University of London, Royal Holloway and Bedford New College, Egham, Surrey TW20 0EX, United Kingdom }
\affiliation{University of Louisville, Louisville, Kentucky 40292, USA }
\affiliation{Ludwig Maximilians University, 80539 Munich, Germany }
\affiliation{Luther College, Decorah, Iowa 52101, USA }
\affiliation{Johannes Gutenberg-Universit\"at Mainz, Institut f\"ur Kernphysik, D-55099 Mainz, Germany }
\affiliation{University of Manchester, Manchester M13 9PL, United Kingdom }
\affiliation{University of Maribor, 2000 Maribor, Slovenia }
\affiliation{University of Maryland, College Park, Maryland 20742, USA }
\affiliation{Massachusetts Institute of Technology, Laboratory for Nuclear Science, Cambridge, Massachusetts 02139, USA }
\affiliation{Max-Planck-Institut f\"ur Physik, 80805 M\"unchen, Germany }
\affiliation{McGill University, Montr\'eal, Qu\'ebec, Canada H3A 2T8 }
\affiliation{School of Physics, University of Melbourne, Victoria 3010, Australia }
\affiliation{INFN Sezione di Milano$^{\mathrm{a}}$; Dipartimento di Fisica, Universit\`a di Milano$^{\mathrm{b}}$, I-20133 Milano, Italy }
\affiliation{University of Mississippi, University, Mississippi 38677, USA }
\affiliation{Universit\'e de Montr\'eal, Physique des Particules, Montr\'eal, Qu\'ebec, Canada H3C 3J7  }
\affiliation{Moscow Physical Engineering Institute, Moscow 115409, Russian Federation }
\affiliation{Moscow Institute of Physics and Technology, Moscow Region 141700, Russian Federation }
\affiliation{Graduate School of Science, Nagoya University, Nagoya 464-8602, Japan }
\affiliation{Kobayashi-Maskawa Institute, Nagoya University, Nagoya 464-8602, Japan }
\affiliation{INFN Sezione di Napoli$^{\mathrm{a}}$; Dipartimento di Scienze Fisiche, Universit\`a di Napoli Federico II$^{\mathrm{b}}$, I-80126 Napoli, Italy }
\affiliation{Nara Women's University, Nara 630-8506, Japan }
\affiliation{National Central University, Chung-li 32054, Taiwan }
\affiliation{NIKHEF, National Institute for Nuclear Physics and High Energy Physics, NL-1009 DB Amsterdam, The Netherlands }
\affiliation{National United University, Miao Li 36003, Taiwan }
\affiliation{Department of Physics, National Taiwan University, Taipei 10617, Taiwan }
\affiliation{H. Niewodniczanski Institute of Nuclear Physics, Krakow 31-342, Poland }
\affiliation{Niigata University, Niigata 950-2181, Japan }
\affiliation{University of Notre Dame, Notre Dame, Indiana 46556, USA }
\affiliation{Ohio State University, Columbus, Ohio 43210, USA }
\affiliation{Osaka City University, Osaka 558-8585, Japan }
\affiliation{Pacific Northwest National Laboratory, Richland, Washington 99352, USA }
\affiliation{INFN Sezione di Padova$^{\mathrm{a}}$; Dipartimento di Fisica, Universit\`a di Padova$^{\mathrm{b}}$, I-35131 Padova, Italy }
\affiliation{Laboratoire de Physique Nucl\'eaire et de Hautes Energies, IN2P3/CNRS, Universit\'e Pierre et Marie Curie-Paris6, Universit\'e Denis Diderot-Paris7, F-75252 Paris, France }
\affiliation{Peking University, Beijing 100871, China }
\affiliation{INFN Sezione di Perugia$^{\mathrm{a}}$; Dipartimento di Fisica, Universit\`a di Perugia$^{\mathrm{b}}$, I-06123 Perugia, Italy }
\affiliation{INFN Sezione di Pisa$^{\mathrm{a}}$; Dipartimento di Fisica, Universit\`a di Pisa$^{\mathrm{b}}$; Scuola Normale Superiore di Pisa$^{\mathrm{c}}$, I-56127 Pisa, Italy }
\affiliation{University of Pittsburgh, Pittsburgh, Pennsylvania 15260, USA }
\affiliation{Princeton University, Princeton, New Jersey 08544, USA }
\affiliation{Punjab Agricultural University, Ludhiana 141004, India }
\affiliation{INFN Sezione di Roma$^{\mathrm{a}}$; Dipartimento di Fisica, Universit\`a di Roma La Sapienza$^{\mathrm{b}}$, I-00185 Roma, Italy }
\affiliation{Universit\"at Rostock, D-18051 Rostock, Germany }
\affiliation{Rutherford Appleton Laboratory, Chilton, Didcot, Oxon, OX11 0QX, United Kingdom }
\affiliation{CEA, Irfu, SPP, Centre de Saclay, F-91191 Gif-sur-Yvette, France }
\affiliation{University of Science and Technology of China, Hefei 230026, China }
\affiliation{Soongsil University, Seoul 156-743, South Korea }
\affiliation{SLAC National Accelerator Laboratory, Stanford, California 94309 USA }
\affiliation{University of South Carolina, Columbia, South Carolina 29208, USA }
\affiliation{Southern Methodist University, Dallas, Texas 75275, USA }
\affiliation{Stanford University, Stanford, California 94305-4060, USA }
\affiliation{State University of New York, Albany, New York 12222, USA }
\affiliation{Sungkyunkwan University, Suwon 440-746, South Korea }
\affiliation{School of Physics, University of Sydney, NSW 2006, Australia }
\affiliation{Department of Physics, Faculty of Science, University of Tabuk, Tabuk 71451, Kingdom of Saudi Arabia }
\affiliation{Tata Institute of Fundamental Research, Mumbai 400005, India }
\affiliation{Excellence Cluster Universe, Technische Universit\"at M\"unchen, 85748 Garching}
\affiliation{Tel Aviv University, School of Physics and Astronomy, Tel Aviv, 69978, Israel }
\affiliation{University of Tennessee, Knoxville, Tennessee 37996, USA }
\affiliation{University of Texas at Austin, Austin, Texas 78712, USA }
\affiliation{University of Texas at Dallas, Richardson, Texas 75083, USA }
\affiliation{Toho University, Funabashi 274-8510, Japan }
\affiliation{Tohoku University, Sendai 980-8578, Japan }
\affiliation{Earthquake Research Institute, University of Tokyo, Tokyo 113-0032, Japan }
\affiliation{Department of Physics, University of Tokyo, Tokyo 113-0033, Japan }
\affiliation{Tokyo Institute of Technology, Tokyo 152-8550, Japan }
\affiliation{Tokyo Metropolitan University, Tokyo 192-0397, Japan }
\affiliation{INFN Sezione di Torino$^{\mathrm{a}}$; Dipartimento di Fisica, Universit\`a di Torino$^{\mathrm{b}}$, I-10125 Torino, Italy }
\affiliation{INFN Sezione di Trieste$^{\mathrm{a}}$; Dipartimento di Fisica, Universit\`a di Trieste$^{\mathrm{b}}$, I-34127 Trieste, Italy }
\affiliation{IFIC, Universitat de Valencia-CSIC, E-46071 Valencia, Spain }
\affiliation{University of Victoria, Victoria, British Columbia, Canada V8W 3P6 }
\affiliation{CNP, Virginia Polytechnic Institute and State University, Blacksburg, Virginia 24061, USA }
\affiliation{Department of Physics, University of Warwick, Coventry CV4 7AL, United Kingdom }
\affiliation{Wayne State University, Detroit, Michigan 48202, USA }
\affiliation{University of Wisconsin, Madison, Wisconsin 53706, USA }
\affiliation{Yamagata University, Yamagata 990-8560, Japan }
\affiliation{Yonsei University, Seoul 120-749, South Korea }

\author{A.~Abdesselam\bel}\affiliation{Department of Physics, Faculty of Science, University of Tabuk, Tabuk 71451, Kingdom of Saudi Arabia } 
\author{I.~Adachi\bel}\affiliation{High Energy Accelerator Research Organization (KEK), Tsukuba 305-0801, Japan }\affiliation{SOKENDAI (The Graduate University for Advanced Studies), Hayama 240-0193, Japan } 
\author{A.~Adametz\bbr}\affiliation{Universit\"at Heidelberg, Physikalisches Institut, D-69120 Heidelberg, Germany }
\author{T.~Adye\bbr}\affiliation{Rutherford Appleton Laboratory, Chilton, Didcot, Oxon, OX11 0QX, United Kingdom }
\author{H.~Ahmed\bbr}\affiliation{Physics Department, Jazan University, Jazan 22822, Kingdom of Saudi Arabia }
\author{H.~Aihara\bel}\affiliation{Department of Physics, University of Tokyo, Tokyo 113-0033, Japan } 
\author{S.~Akar\bbr}\affiliation{Laboratoire de Physique Nucl\'eaire et de Hautes Energies, IN2P3/CNRS, Universit\'e Pierre et Marie Curie-Paris6, Universit\'e Denis Diderot-Paris7, F-75252 Paris, France }
\author{M.~S.~Alam\bbr}\affiliation{State University of New York, Albany, New York 12222, USA }
\author{J.~Albert\bbr}\affiliation{University of Victoria, Victoria, British Columbia, Canada V8W 3P6 }
\author{S.~Al~Said\bel}\affiliation{Department of Physics, Faculty of Science, University of Tabuk, Tabuk 71451, Kingdom of Saudi Arabia }\affiliation{Department of Physics, Faculty of Science, King Abdulaziz University, Jeddah 21589, Kingdom of Saudi Arabia } 
% ...

\collaboration{The {\bbr}\babar\ and {\bel}Belle Collaborations}

\begin{abstract}
We report a measurement of the time-dependent \CP asymmetry of $\Bzb \to D^{(*)}_{\CP} h^{0}$ decays, where the light neutral hadron $h^{0}$ is a $\pi^{0}$, $\eta$ or $\omega$ meson,
and the neutral $D$ meson is reconstructed in the \CP eigenstates $K^{+}K^{-}$, $K_{S}^{0}\pi^{0}$ or $K_{S}^{0}\omega$.
The measurement is performed combining the final data samples collected at the $\Upsilon\left(4S\right)$ resonance by the \babar\ and Belle experiments at the asymmetric-energy \B factories PEP-II at SLAC and KEKB at KEK, respectively.
The data samples contain $( 471 \pm 3 )\times 10^6\, \B\Bbar$ pairs recorded by the \babar\ detector and $( 772 \pm 11 )\times 10^6\, \B\Bbar$ pairs recorded by the Belle detector.
We measure the \CP asymmetry parameters $-\eta_{f}\mathcal{S} = +0.66 \pm 0.10\,(\rm{stat.}) \pm 0.06\,(\rm{syst.})$ and $\mathcal{C} = -0.02 \pm 0.07\,(\rm{stat.}) \pm 0.03\,(\rm{syst.})$.
These results correspond to the first observation of \CP violation in $\Bzb \to D^{(*)}_{\CP} h^{0}$ decays. The hypothesis of no mixing-induced \CP violation is excluded in these decays at the level of $5.4$ standard deviations.
\end{abstract}


\maketitle

\tighten

{\renewcommand{\thefootnote}{\fnsymbol{footnote}}}
\setcounter{footnote}{0}

In the standard model (SM) of electroweak interactions, \CP violation arises from an irreducible complex phase in the three-family Cabibbo-Kobayashi-Maskawa (CKM) quark-mixing matrix~\cite{CabibboKobayashiMaskawa}.
The \babar\ and Belle experiments have established \CP violating effects in the \B meson system~\cite{CPV_observation_BaBar,CPV_observation_Belle,directCPV_BaBar,directCPV_Belle}.
Both experiments use their measurements of the mixing-induced \CP violation in $b \to c\bar{c}s$ transitions~\cite{BaBar_btoccs,Belle_btoccs} to determine precisely the parameter $\sin(2\beta) \equiv \sin(2\phi_1)$
[\babar\ uses $\beta$ and Belle uses $\phi_1$, hereinafter $\beta$ is used].
The angle $\beta$ is defined as $\arg\left[-V^{}_{cd} V^{*}_{cb} / V^{}_{td} V^{*}_{tb} \right]$, where $V_{ij}$ is the CKM matrix element of quarks $i$, $j$.

A complementary and theoretically clean approach to access $\beta$ is provided by $\Bzb \to D^{(*)0} h^{0}$ decays, where $h^{0} \in \{\pi^{0}, \eta, \omega \}$ denotes a light neutral hadron.
These decays are dominated by CKM-favored $b \to c\bar{u}d$ tree amplitudes.
CKM-disfavored $b \to u\bar{c}d$ amplitudes carrying different weak phases contribute also to the decays, but are suppressed by $V^{}_{ub} V^{*}_{cd} / V^{}_{cb} V^{*}_{ud} \approx 0.02$ relative to the leading amplitudes.
An interference between the decay amplitudes without and with \Bz-\Bzb mixing emerges if the neutral $D$ meson decays to a \CP eigenstate $D_{\CP}$.
Neglecting the suppressed amplitudes, the time evolution of $\Bzb \to D^{(*)}_{\CP} h^{0}$ decays is governed by $\beta$~\cite{Fleischer2003}.
Because only tree-level amplitudes contribute to $\Bzb \to D^{(*)0} h^{0}$ decays, these decays are not sensitive to most models of physics beyond the standard model (BSM).
However, the measurement of the time-dependent \CP violation enables testing of the measurements of $b \to c\bar{c}s$ transitions~\cite{BaBar_btoccs,Belle_btoccs}
and provides a SM reference for the BSM searches in the mixing-induced \CP violation of $b \to s$ penguin-mediated \B meson decays~\cite{MeasurementsbtoqqbarsBaBar1,MeasurementsbtoqqbarsBelle0,MeasurementsbtoqqbarsBaBar0,MeasurementsbtoqqbarsBelle1}.
Any sizable deviation in the \CP asymmetry of $\Bzb \to D^{(*)}_{\CP} h^{0}$ decays from processes involving $b \to c\bar{c}s$ or penguin-mediated $b \to s$ transitions would point to BSM.
Such deviations could for example be caused by unobserved heavy particles contributing to loop diagrams in $b \to c\bar{c}s$ or $b \to s$ penguin transitions~\cite{newphysicsinloops}.

An experimental difficulty in the use of $\Bzb \to D^{(*)}_{\CP} h^{0}$ decays arises from low \B and $D$ meson branching fractions [$\mathcal{O}(10^{-4})$ and $\mathcal{O}(\leq10^{-2})$, respectively] and low reconstruction efficiencies.
Previous measurements performed separately by the \babar\ and Belle collaborations were not able to establish \CP violation in these or related decays~\cite{Belle_D0h0_2006,BaBar_D0h0_2007_twobody,BaBar_D0h0_2007_threebody}.

In this Letter, we present a measurement of the time-dependent \CP violation in $\Bzb \to D^{(*)}_{\CP} h^{0}$ decays.
For the first time, we combine the large final data samples collected by the \babar\ and Belle experiments.
This new approach enables time-dependent \CP violation measurements in the neutral \B meson system with unprecedented sensitivity.

The time-dependent rate of a neutral \B meson decaying to a \CP eigenstate is given by
\begin{align}
  g ( \Delta t ) = 
  \frac{ e^{ - \lvert \Delta t \rvert / \tau_{\Bz} } }{ 4 \tau_{\Bz} } &
  \lbrace
  1 + q [
  \mathcal{S} \sin( \Delta m_{d} \Delta t ) \nonumber \\
  & - \mathcal{C}  \cos( \Delta m_{d} \Delta t )
  ] \rbrace , \label{decay_rate}
\end{align}
where $q = +1 \,(-1)$ represents the $b$-flavor content when the accompanying \B meson is tagged as a \Bz (\Bzb) and
$\Delta t$ denotes the proper time interval between the decays of the two \B mesons produced in an $\Upsilon\left(4S\right)$ decay.
The neutral \B meson lifetime is represented by $\tau_{\Bz}$, and the \Bz-\Bzb mixing frequency by $\Delta m_{d}$.
Neglecting the CKM-disfavored decay amplitudes in $\Bzb \to D^{(*)}_{\CP} h^{0}$ decays, the SM predicts $\mathcal{S} = -\eta_{f} \sin(2\beta)$ and $\mathcal{C} = 0$,
where $\eta_{f}$ is the \CP eigenvalue of the final state, and $\mathcal{S}$ and $\mathcal{C}$, respectively, quantify mixing-induced and direct \CP violation~\cite{directCPV_A_or_C}.

This analysis is based on data samples collected at the $\Upsilon(4S)$ resonance containing
$( 471 \pm 3 )\times 10^6\, \B\Bbar$ pairs recorded with the \babar\ detector at the PEP-II asymmetric-energy $e^+ e^-$ (3.1 on 9~GeV) collider~\cite{PEPII} and
$( 772 \pm 11 )\times 10^6\, \B\Bbar$ pairs recorded with the Belle detector at the KEKB asymmetric-energy $e^+ e^-$ (3.5 on 8~GeV) collider~\cite{KEKB}.
At \babar\ (Belle) the $\Upsilon(4S)$ is produced with a Lorentz boost of $\beta\gamma = 0.560$ ($0.425$),
allowing the measurement of $\Delta t$ from the displacement of the decay vertices of the two \B mesons.
The \babar\ and Belle detectors are described in Refs.~\cite{BaBarDetector,BelleDetector}.

Reconstructed tracks of charged particles are considered as kaon and pion candidates. Kaons are identified using the particle identification techniques described in Refs.~\cite{BaBarDetector,BelleDetector}.
Photons are reconstructed from energy deposits in the electromagnetic calorimeters, and the energy of photon candidates is required to be at least $30\,\textrm{MeV}$.
Combinations of two photons are considered as $\pi^{0}$ meson candidates if the reconstructed invariant mass is between $115$ and $150~\textrm{MeV/}c^2$.
Candidate $\eta$ mesons are reconstructed in the decay modes $\eta \to \gamma\gamma$ and $\pi^{+}\pi^{-}\pi^{0}$.
The invariant mass is required to be within $20~\textrm{MeV/}c^2$ of the nominal mass~\cite{PDG} for $\eta \to \gamma\gamma$ candidates, and within $10~\textrm{MeV/}c^2$ for $\eta \to \pi^{+}\pi^{-}\pi^{0}$ candidates.
For each photon in the $\eta \to \gamma\gamma$ decay mode a minimal energy of $50\,\textrm{MeV}$ is required.

For $\omega$ mesons the decay mode $\omega \to \pi^{+}\pi^{-}\pi^{0}$ is reconstructed with invariant mass required to be within $15~\textrm{MeV/}c^2$ of the nominal mass~\cite{PDG}.
Neutral kaons are reconstructed in the decay mode $K_{S}^{0} \to \pi^{+}\pi^{-}$, with invariant mass required to be within $15~\textrm{MeV/}c^2$ of the nominal mass~\cite{PDG}.
The requirements exploiting the $K_{S}^{0}$ decay vertex displacement from the interaction point (IP) described in Refs.~\cite{BELLEKshortselection,BaBar_D0h0_2007_twobody} are applied.
Neutral $D$ mesons are reconstructed in the decay modes to \CP eigenstates $D_{\CP} \to K^{+}K^{-}$, $K_{S}^{0}\pi^{0}$ and $K_{S}^{0}\omega$.
The invariant mass is required to be within $12~\textrm{MeV/}c^2$ of the nominal mass~\cite{PDG} for $D_{\CP} \to K^{+}K^{-}$ candidates,
within $25~\textrm{MeV/}c^2$ for $D_{\CP} \to K_{S}^{0}\pi^{0}$ candidates,
and within $20~\textrm{MeV/}c^2$ for $D_{\CP} \to K_{S}^{0}\omega$ candidates.
We reconstruct $D^{*0}$ mesons in the decay mode $D^{*0} \to \Dz\pi^{0}$, and the invariant mass must be within $3~\textrm{MeV/}c^2$ of the nominal mass~\cite{PDG}.

Neutral \B mesons are reconstructed in the \CP-even ($\eta_{f} = +1$) final states
$\Bzb \to D_{\CP} \pi^{0}$ and $D_{\CP} \eta$ (with $D_{\CP} \to K_{S}^{0}\pi^{0}$, $K_{S}^{0}\omega$),
$\Bzb \to D_{\CP} \omega$ (with $D_{\CP} \to K_{S}^{0}\pi^{0}$),
$\Bzb \to D_{\CP}^{*} \pi^{0}$ and $D_{\CP}^{*} \eta$ (with $D_{\CP} \to K^{+}K^{-}$),
and in the \CP-odd ($\eta_{f} = -1$) final states
$\Bzb \to D_{\CP} \pi^{0}$, $D_{\CP} \eta$, $D_{\CP} \omega$ (with $D_{\CP} \to K^{+}K^{-}$),
and $\Bzb \to D_{\CP}^{*} \pi^{0}$ and $D_{\CP}^{*} \eta$ (with $D_{\CP} \to K_{S}^{0}\pi^{0}$)~\cite{CC}.

Neutral \B mesons are selected by the beam-energy-constrained mass $M_{\rm bc} \equiv m_{\rm ES} = \sqrt{ ( E^{*}_{\rm beam} / c^{2} )^{2} - ( p^{*}_{\B} / c )^{2} }$ [\babar\ uses $m_{\rm ES}$ and Belle uses $M_{\rm bc}$, hereinafter $M_{\rm bc}$ is used]
and by the energy difference $\Delta E = E^{*}_{\B} - E^{*}_{\rm beam}$, where $E^{*}_{\rm beam}$ denotes the energy of the beam, and $p^{*}_{\B}$ and $E^{*}_{\B}$ are the momentum and energy of the \B meson candidates, evaluated in the $e^+ e^-$ center-of-mass (c.m.) frame.
The selected regions are $5.2~\mathrm{GeV}/c^{2} < M_{\rm bc} < 5.3~\mathrm{GeV}/c^{2}$
and $-100\,\mathrm{MeV} < \Delta E < 100\,\mathrm{MeV}$, except for $\Bzb \to D_{\CP}^{(*)} \pi^{0}$ decays, where $-75\,\mathrm{MeV} < \Delta E < 100\,\mathrm{MeV}$ is required to exclude tails from partially reconstructed
$\Bm \to D^{(*)0} \rho^{-}$ decays peaking at $\Delta E \approx -250\,\mathrm{MeV}$.

In $\Bzb \to D^{0} \omega$ and in $D^{0} \to K_{S}^{0}\omega$ decays, the $\omega$ vector mesons are polarized.
The angular distribution of $\omega \to \pi^{+}\pi^{-}\pi^{0}$ decays is exploited to discriminate against background.
The quantity $\cos\theta_{N}$ is defined as the cosine of the angle between the neutral \B meson direction and the normal to the $\pi^{+}\pi^{-}\pi^{0}$ plane in the $\omega$ meson rest frame.
A requirement of $\lvert \cos\theta_{N} \rvert > 0.3$ is applied.

After applying the above selection requirements, the average multiplicity of reconstructed $\Bzb \to D^{(*)}_{\CP} h^{0}$ candidates in an event is 1.3.
In case of multiple \B meson candidates in an event, one candidate is selected using a criterion based on the deviations of the reconstructed $D^{(*)}$ and $h^{0}$ meson masses from the nominal values.
The probability for this method to select the correct signal is $82\%$ ($81\%$) for \babar\ (Belle).

In $\Bzb \to D^{(*)}_{\CP} h^{0}$ decays, the dominant source of background originates from $e^{+}e^{-} \to q \overline{q}$  $( q \in \{ u, d, s, c \} )$ continuum events.
This background is suppressed by using neural network (NN) multivariate classifiers that combine information characterizing the shape of an event~\cite{Neurobayes}.
The observables included in the NNs are the ratio $R_{2}$ of the second to the zeroth order Fox-Wolfram moment, a combination of 16 modified Fox-Wolfram moments~\cite{FWmoments},
the sphericity of the event~\cite{Sphericity}, and $\cos\theta_{\B}^{*}$, where $\theta_{\B}^{*}$ is the angle between the direction of the reconstructed \B meson and the beam direction in the c.m.\ frame.
The NN selection reduces the background by $89.3\%$ ($91.8\%$) and has a signal efficiency of $75.5\%$ ($74.3\%$) for \babar\ (Belle).

\begin{figure}[htb]
\includegraphics[width=0.475\textwidth]{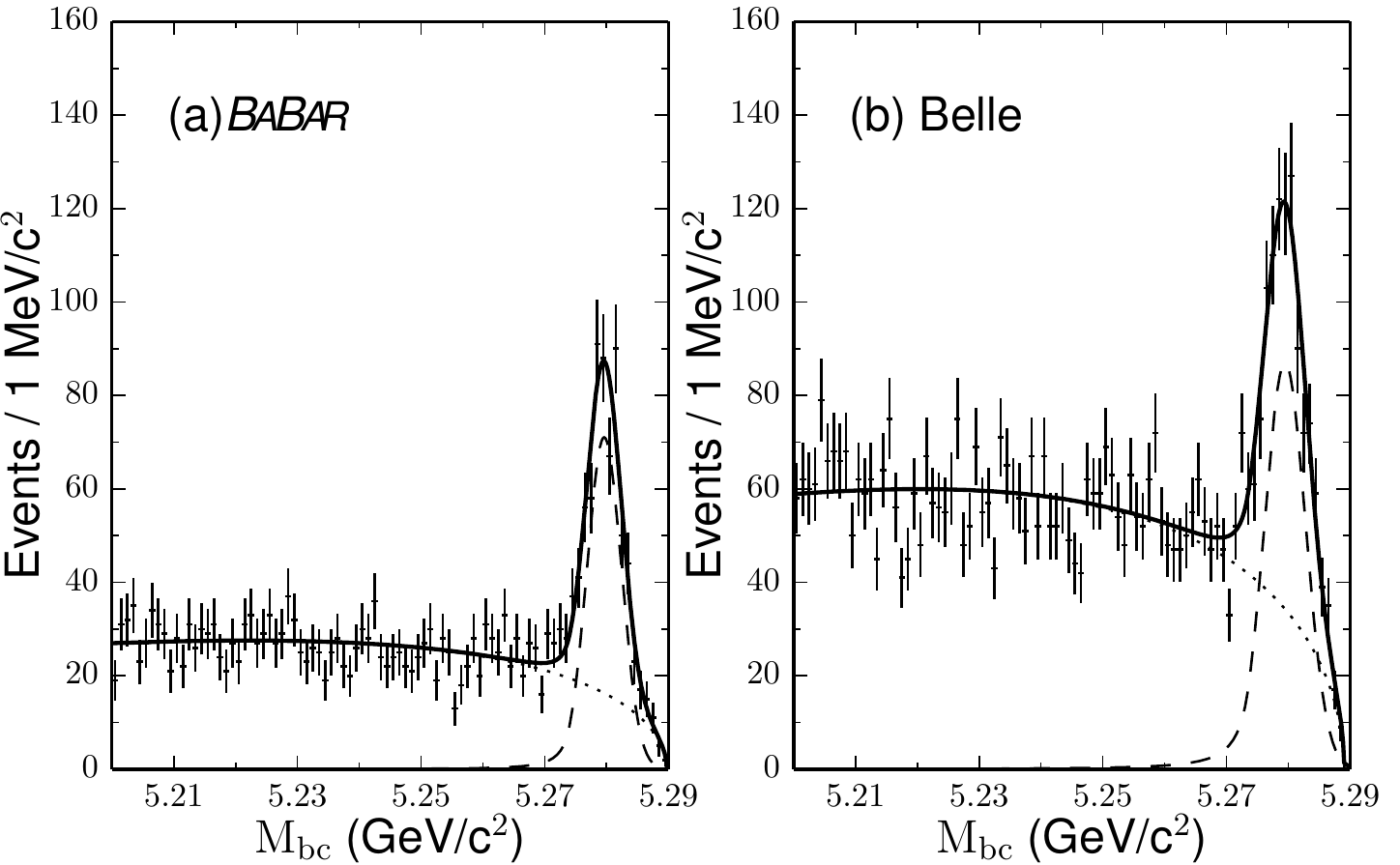}
\caption{
The $M_{\rm bc}$ distributions (data points with error bars) and fit projections (solid line)
of $\Bzb \to D^{(*)}_{\CP} h^{0}$ decays for (a) \babar\ and (b) Belle.
The dashed (dotted) lines represent projections of the signal (background) fit components.
}
\label{figure_mbc_deltae_distributions}
\end{figure}

The signal yields are determined by unbinned maximum likelihood fits to the $M_{\rm bc}$ distributions.
In the fits, the signal component is parametrized by a Crystal Ball function~\cite{CBfunction} and the background component is modeled by an ARGUS function~\cite{ARGUSfunction}.
The experimental $M_{\rm bc}$ distributions and fit projections are shown in Fig.~\ref{figure_mbc_deltae_distributions}.
The signal yields are summarized in Table~\ref{table:data_yields_decay_modes}.

\begin{table}[htb]
\caption{Summary of $\Bzb \to D^{(*)}_{\CP} h^{0}$ signal yields.}
\label{table:data_yields_decay_modes}
\begin{tabular}
{@{\hspace{0.5cm}}l@{\hspace{0.5cm}}  @{\hspace{0.5cm}}c@{\hspace{0.5cm}}  @{\hspace{0.5cm}}c@{\hspace{0.5cm}}}
\hline \hline
Decay mode & \babar & Belle \\
\hline
$\Bzb \to D^{}_{\CP} \pi^{0}$         & $ 241 \pm 22 $ & $ 345 \pm 25 $ \\
$\Bzb \to D^{}_{\CP} \eta$            & $ 106 \pm 14 $ & $ 148 \pm 18 $ \\
$\Bzb \to D^{}_{\CP} \omega$          & $  66 \pm 10 $ & $ 151 \pm 17 $ \\
$\Bzb \to D^{*}_{\CP} \pi^{0}$        & $  72 \pm 12 $ & $  80 \pm 14 $ \\
$\Bzb \to D^{*}_{\CP} \eta$           & $  39 \pm  8 $ & $  39 \pm 10 $ \\
\hline
$\Bzb \to D^{(*)}_{\CP} h^{0}$ total  &  $508 \pm 31$  & $ 757 \pm 44 $ \\
\hline \hline
\end{tabular}
\end{table}

The time-dependent \CP violation measurement is performed using established \babar\ and Belle techniques for the vertex reconstruction, the flavor-tagging, and the modeling of $\Delta t$ resolution effects
(see Refs.~\cite{BaBar_btoccs,Belle_btoccs,BaBar_btoccs2002,BelleVertexResolution,BelleTaggingNIM,BFactoriesBook}) and is briefly summarized below.
The proper time interval $\Delta t$ is given as $\frac{\Delta z}{\mathrm{c} \beta \gamma}$, where $\Delta z$ is the distance between the decay vertices of the signal \B meson and of the accompanying \B meson.
The $\Bzb \to D^{(*)}_{\CP} h^{0}$ signal decay vertex is reconstructed by a kinematic fit including information about the IP position.
For Belle, an iterative hierarchical vertex reconstruction algorithm following a bottom-up approach starting with the final state particles is applied, while for \babar\ the vertex reconstruction includes simultaneously the complete \B meson decay tree including all secondary decays.
In the kinematic fits, the invariant masses of $\pi^{0}$, $\eta$, $\omega$, and $D_{\CP}$ candidates are constrained to their nominal values~\cite{PDG}.
The decay vertex and the $b$-flavor content of the accompanying \B meson are estimated from reconstructed decay products not assigned to the signal \B meson.
The $b$-flavor content is inferred by flavor-tagging procedures described in Refs.~\cite{BaBar_btoccs,BelleTaggingNIM}. The applied algorithms account for different signatures such as the presence and properties of prompt leptons, charged kaons and pions
originating from the decay of the accompanying \B meson, and assign a flavor and an associated probability.
Selection requirements on the quality of the reconstructed decay vertices and the $\Delta t$ measurements are applied.

The \CP violation measurement is performed by maximizing the log-likelihood function
\begin{equation}
\ln \mathcal{L} = \sum \limits_{i} \ln \mathcal{P}_{i}^{\mathrm{\scriptsize\babar}} + \sum \limits_{j} \ln \mathcal{P}_{j}^{\mathrm{Belle}}, \label{equation_loglikelihood}
\end{equation}
where the indices $i$ and $j$ denote the events reconstructed from \babar\ and Belle data, respectively.
The probability density function (p.d.f.) describing the $\Delta t$ distribution for \babar\ is defined by
\begin{align}
\mathcal{P}^{\mathrm{\scriptsize\babar}} =
\sum\limits_{k}
f_{k} 
\int
\left[
{P}_{k} \left( \Delta t' \right)
R_{k} \left( \Delta t - \Delta t' \right) 
\right]
d\left( \Delta t' \right ) ,
\end{align}
and for Belle by
\begin{align}
\mathcal{P}^{\mathrm{Belle}} =
 ( 1 & - f_{\rm ol} )
\sum\limits_{k}
f_{k} 
\int
\left[
{P}_{k} \left( \Delta t' \right)
R_{k} \left( \Delta t - \Delta t' \right) 
\right]
d\left( \Delta t' \right ) \nonumber \\
+ & f_{\rm ol} P_{\rm ol} \left( \Delta t \right) ,
\end{align}
where the index $k$ represents the signal and background p.d.f. components.
The symbol ${P}_{k}$ denotes the p.d.f. describing the proper time interval of the particular physical process, and $R_{k}$ refers to the corresponding resolution function.
The fractions $f_{k}$ are evaluated on an event-by-event basis as a function of $M_{\rm bc}$.
Belle treats outlier events with large $\Delta t$ using a broad Gaussian function in the p.d.f. component $P_{\rm ol}$ with a small fraction of $f_{\rm ol} \approx 2 \times 10^{-4}$, while \babar\ includes outlier effects in the resolution function.
The signal p.d.f. is constructed from the decay rate in Eq.~(\ref{decay_rate}), including the effect of incorrect flavor assignments
and convolution with resolution functions to account for the finite vertex resolution.
The models of the $\Delta t$ resolution effects at \babar\ and Belle follow different empirical approaches and are described in detail in Refs.~\cite{BaBar_btoccs,BelleVertexResolution}.
The background p.d.f.s for \babar\ and Belle are
composed of the sum of a Dirac delta function to model prompt background decays and an exponential p.d.f. for decays with effective lifetimes. The background p.d.f. is convolved with a resolution function modeled as the sum of two Gaussian functions.
The background parameters are fixed to values obtained by fits to the events in the $M_{\rm bc} < 5.26~\mathrm{GeV}/c^{2}$ sidebands.

\begin{figure}[htb]
\includegraphics[width=0.475\textwidth]{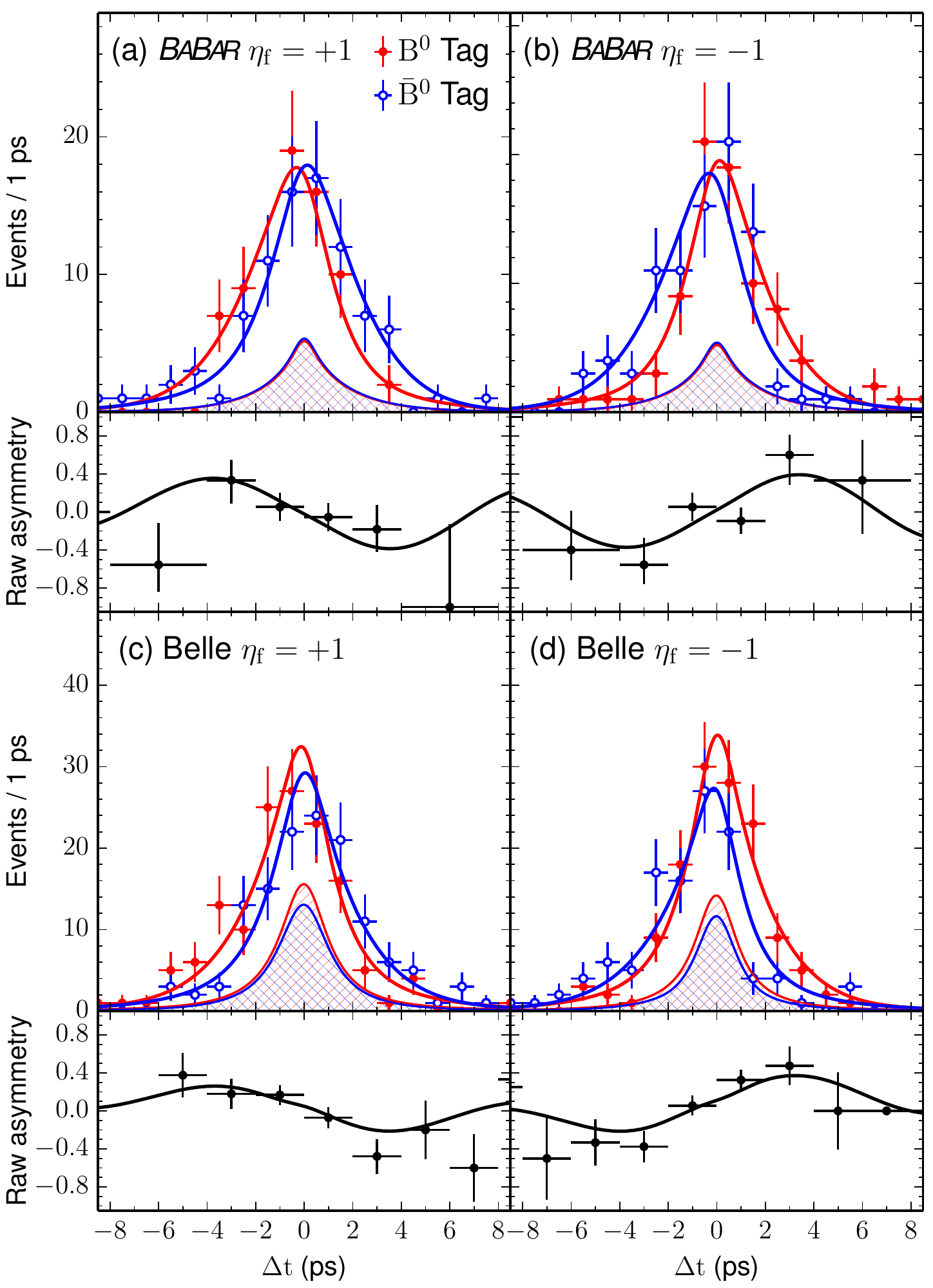}
\caption{
(color online). The proper time interval distributions (data points with error bars) for $\Bz$ tags (red) and $\Bzb$ tags (blue) and the \CP asymmetries of $\Bzb \to D^{(*)}_{\CP} h^{0}$ decays for (a)-(b) \babar\ and (c)-(d) Belle for candidates
associated with high quality flavor tags.
The solid lines show projections of the sum of signal and background components in the fit, while the hatched areas show only the background components.
}
\label{figure_cpfit}
\end{figure}

The combined \babar\ and Belle measurement is performed by maximizing Eq.~(\ref{equation_loglikelihood}) for events in the $5.27~\mathrm{GeV}/c^{2} < M_{\rm bc} < 5.29~\mathrm{GeV}/c^{2}$ signal region.
The values of $\tau_{\Bz}$ and $\Delta m_{d}$ are fixed to the world averages~\cite{PDG}.
The free parameters in the fit are $\mathcal{S}$ and $\mathcal{C}$.
The result is
\begin{alignat}{2}
-\eta_{f}\mathcal{S} &= +0.66 \pm 0.10\,(\rm{stat.}) &&\pm 0.06\,(\rm{syst.}), \nonumber \\
\mathcal{C} &= -0.02 \pm 0.07\,(\rm{stat.}) &&\pm 0.03\,(\rm{syst.}).
\end{alignat}
The linear correlation between $-\eta_{f}\mathcal{S}$ and $\mathcal{C}$ is $-4.9\%$.
Through comparison of the log-likelihood of the fit to the distribution from an ensemble test performed with input from the data distributions, a $p$-value of $0.46$ is obtained.
The flavor-tagged proper time interval distributions and projections of the fit are shown in Fig.~\ref{figure_cpfit}.

The evaluation of the systematic uncertainties in the \CP violation parameters follows standard approaches of the \babar\ and Belle experiments described in detail in Refs.~\cite{BaBar_btoccs,Belle_btoccs,BFactoriesBook};
the results are summarized in Table~\ref{systemtic_uncertainties_cpfit}.
For the vertex reconstruction, the sources of systematic uncertainties include the applied constraints and selection requirements on the vertex fits of the signal \B meson and the accompanying \B meson, and on the $\Delta t$ fit range.
These contributions are estimated by variations of the constraints and selection requirements.
The systematic uncertainties due to the misalignment of the silicon vertex detectors are estimated by Monte Carlo (MC) simulations.
For \babar, the uncertainty of the $z$ scale is estimated by variations of the $z$ scale and corresponding uncertainties. For Belle, a possible $\Delta t$ bias is estimated using MC simulations.
The systematic uncertainties due to the $\Delta t$ resolution functions, the parameterization of the $\Delta t$ background p.d.f., the calculation of the signal purity, the flavor-tagging, and the physics parameters $\tau_{\Bz}$ and $\Delta m_{d}$ are
estimated by variation of the fixed parameters within their uncertainties.
Fit biases are estimated using large samples of MC-simulated signal decays.
The contribution of backgrounds that have the same final states as the reconstructed $\Bzb \to D^{(*)}_{\CP} h^{0}$ decay modes and that can peak in the $M_{\rm bc}$ signal region
is estimated using $D$ meson mass sidebands on data and using generic $\B\Bbar$ MC samples. These backgrounds account for less than $8\%$ of the signal and consist mainly of flavor-specific decays such as partially reconstructed $\Bm \to D^{(*)0} \rho^{-}$ decays.
The systematic uncertainty due to this peaking background is estimated using MC simulations in which the peaking background is modeled, and the nominal fit procedure, which neglects this peaking background, is applied.
The effect of interference between $b \to c\bar{u}d$ and $\bar{b} \to \bar{u}c\bar{d}$ decay amplitudes of the accompanying \B meson is estimated using MC simulations that account for possible deviations from the time evolution described by Eq.~(\ref{decay_rate})~\cite{TagSideInterference}.
Possible correlations between \babar\ and Belle are accounted for in the evaluation of the contributions due to the physics parameters, the peaking background, and the tag-side interference.
In the MC studies described above, the largest deviations are assigned as systematic uncertainties. The total systematic uncertainty is the quadratic sum of all contributions.

\begin{table}[htb]
\caption{Summary of systematic uncertainties for the time-dependent \CP violation measurement in $\Bzb \to D^{(*)}_{\CP} h^{0}$ decays (in units of $10^{-2}$).}
\label{systemtic_uncertainties_cpfit}
\begin{tabular}
 {@{\hspace{0.5cm}}l@{\hspace{0.5cm}}  @{\hspace{0.5cm}}c@{\hspace{0.5cm}}  @{\hspace{0.5cm}}c@{\hspace{0.5cm}}}
\hline \hline
Source & $\mathcal{S}$ & $\mathcal{C}$ \\
\hline
Vertex reconstruction                                 &   $ 1.5 $    &   $ 1.4 $    \\
$\Delta t$ resolution functions                       &   $ 2.0 $    &   $ 0.4 $    \\
Background $\Delta t$ PDFs                            &   $ 0.4 $    &   $ 0.1 $    \\
Signal purity                                         &   $ 0.6 $    &   $ 0.3 $    \\
Flavor-tagging                                        &   $ 0.3 $    &   $ 0.3 $    \\
Physics parameters                                    &   $ 0.2 $    &   $ <0.1$    \\
Possible fit bias                                     &   $ 0.6 $    &   $ 0.8 $    \\
Peaking background                                    &   $ 4.9 $    &   $ 0.9 $    \\
Tag-side interference                                 &   $ 0.1 $    &   $ 1.4 $    \\
\hline
Total                                                 &   $ 5.6 $    &   $ 2.5 $    \\
\hline \hline
\end{tabular}
\end{table}

The statistical significance of the results is estimated using a likelihood-ratio approach by computing the change in $2\ln \mathcal{L}$ when the \CP violation parameters are fixed to zero.
The effect of systematic uncertainties is included by convolution of the likelihood distributions.
No significant direct \CP violation is observed.
The measurement excludes the hypothesis of no mixing-induced \CP violation in $\Bzb \to D^{(*)}_{\CP} h^{0}$ decays at a confidence level of $1 - 6.6 \times 10^{-8}$, corresponding to a significance of $5.4$ standard deviations.

The analysis is validated by a variety of cross-checks. The same measurement is performed for $\Bzb \to D^{(*)0} h^{0}$ decays with the CKM-favored $D^{0} \to K^{-}\pi^{+}$ decay mode. These decays provide a kinematically similar, high statistics control sample.
The result agrees with the assumption of negligible \CP violation for these decays.
Measurements of the neutral \B meson lifetime using the control sample and $\Bzb \to D^{(*)}_{\CP} h^{0}$ decays yield $\tau_{\Bz} = 1.518 \pm 0.026\,(\rm{stat.})~\mathrm{ps}$
and $\tau_{\Bz} = 1.520 \pm 0.064\,(\rm{stat.})~\mathrm{ps}$, respectively, in agreement with the world average $\tau_{\Bz} = 1.519 \pm 0.005~\mathrm{ps}$~\cite{PDG}.
All measurements for the control sample and for $\Bzb \to D^{(*)}_{\CP} h^{0}$ decays have also been performed for data separated by experiment and by decay mode, and yield consistent results.
The results for $\Bzb \to D^{(*)}_{\CP} h^{0}$ decays separated by experiment are $\sin(2\beta) = 0.52 \pm 0.15\,(\rm{stat.})$ for \babar\ and $0.83 \pm 0.15\,(\rm{stat.})$ for Belle,
and the results separated by the \CP content of the final states are $\sin(2\beta) = 0.52 \pm 0.15\,(\rm{stat.})$ for \CP-even and $0.80 \pm 0.15\,(\rm{stat.})$ for \CP-odd.

In summary, we combine the final \babar\ and Belle data samples, totaling more than $1\,\mathrm{ab}^{-1}$ collected at the $\Upsilon\left(4S\right)$ resonance~\cite{KEKB,BaBarluminosity},
and perform a simultaneous analysis of the data collected by both experiments.
We observe for the first time \CP violation in $\Bzb \to D^{(*)}_{\CP} h^{0}$ decays driven by mixing-induced \CP violation. We measure $\sin(2\beta) = 0.66 \pm 0.10\,(\rm{stat.}) \pm 0.06\,(\rm{syst.})$.
This result agrees within $0.2$ standard deviations with the world average of $\sin(2\beta) = 0.68 \pm 0.02$~\cite{HFAG} measured from $b \to c\bar{c}s$ transitions,
and is consistent with the measurements of $b \to s$ penguin-mediated \B meson decays~\cite{MeasurementsbtoqqbarsBaBar1,MeasurementsbtoqqbarsBelle0,MeasurementsbtoqqbarsBaBar0,MeasurementsbtoqqbarsBelle1}
at current precision.
The presented measurement supersedes the previous \babar\ result for $\Bzb \to D^{(*)}_{\CP} h^{0}$ decays~\cite{BaBar_D0h0_2007_twobody}.

We thank the \pep2\ and KEKB groups for the excellent operation of the
accelerators, and the computing organizations that support \babar\ and Belle.
The Belle experiment wishes to acknowledge the KEK cryogenics group for efficient solenoid operations.
This work was supported by
ARC and DIISR (Australia); FWF (Austria);
NSERC (Canada);
NSFC (China); MSMT (Czechia);
CEA and CNRS-IN2P3 (France);
BMBF, CZF, DFG, and VS (Germany); DST (India); INFN (Italy);
MEXT, JSPS and Nagoya's TLPRC (Japan);
MOE, MSIP, NRF, GSDC of KISTI, and BK21Plus (Korea);
FOM (The Netherlands); NFR (Norway);
MNiSW and NCN (Poland); MES and RFAAE (Russian Federation); ARRS (Slovenia);
IKERBASQUE, MINECO and UPV/EHU (Spain); 
SNSF (Switzerland); NSC and MOE (Taiwan);
STFC (United Kingdom);
BSF (USA-Israel);
and DOE and NSF (USA).
Individuals have received support from the
Marie Curie EIF (European Union)
and the A.~P.~Sloan Foundation (USA).


\begin{thebibliography}{99}
\bibitem{directCPV_A_or_C} Another naming convention for direct \CP asymmetries, $\mathcal{A}$ ($= -\mathcal{C}$), in decays to \CP eigenstates is
also used in the literature.
\bibitem{CC}
In this Letter the inclusion of charge-conjugated decay modes
is implied unless otherwise stated.
\end{thebibliography}
\end{document}